\documentclass[journal,twoside,web]{ieeecolor}
\usepackage{generic}
\usepackage{amsmath,amssymb,amsfonts}
\usepackage{algorithmic}
\usepackage{graphicx}
\usepackage{algorithm,algorithmic}
\usepackage[bookmarks=false]{hyperref}
\usepackage{textcomp}
\usepackage[capitalise]{cleveref}
\usepackage{placeins}
\usepackage{subfig}
\usepackage{stfloats}

\def\BibTeX{{\rm B\kern-.05em{\sc i\kern-.025em b}\kern-.08em
    T\kern-.1667em\lower.7ex\hbox{E}\kern-.125emX}}
    
\usepackage[
    backend=biber, 
    style=ieee,
    sorting=none
]{biblatex}

\begin{document}
\title{DRStageNet: Deep Learning for Diabetic Retinopathy Staging from Fundus Images}

\author{Yevgeniy Men, Jonathan Fhima, Leo Anthony Celi, Lucas Zago Ribeiro, Luis Filipe Nakayama and Joachim A. Behar, \IEEEmembership{Senior Member, IEEE}
\thanks{YM and JB acknowledge the support of the Technion EVPR Fund: Irving \& Branna Sisenwein Research Fund. The research was supported by a cloud computing grant from the Israel Council of Higher Education, administered by the Israel Data Science Initiative. LAC is funded by the National Institute of Health through NIBIB R01 EB017205.}
\thanks{We would like to acknowledge the assistance of ChatGPT, an AI-based language model developed by OpenAI, for its help in editing the English language of this manuscript.}
\thanks{YM is with the Andrew and Erna Viterbi Faculty of Electrical \& Computer Engineering and the Faculty of Biomedical Engineering, Technion, Israel Institute of Technology, Haifa, 3200003, Israel. JF is with the Department of Applied Mathematics and the Biomedical Engineering Faculty, Technion, Israel Institute of Technology, Haifa, 3200003, Israel. LAC is with the Laboratory for Computational Physiology, Massachusetts Institute of Technology, Cambridge, MA 02139, the Division of Pulmonary, Critical Care and Sleep Medicine, Beth Israel Deaconess Medical Center, Boston, MA 02215 and the Department of Biostatistics, Harvard T.H. Chan School of Public Health, Boston, MA 02115. LZR is with the Ophthalmology department, São Paulo Federal University, São Paulo, Brazil. LFN is with the Institute for Medical Engineering and Science, Massachusetts Institute of Technology, Cambridge, MA, USA and the Department of Ophthalmology, São Paulo Federal University, São Paulo, Brazil. JAB is with the Faculty of Biomedical Engineering, Technion, Israel Institute of Technology, Haifa, 3200003, Israel (e-mail: jbehar@technion.ac.il).}
}
\maketitle

%%%%%%%%%%%%%%%%%%%% 
\begin{abstract}
Diabetic retinopathy (DR) is a prevalent complication of diabetes associated with a significant risk of vision loss. Timely identification is critical to curb vision impairment. Algorithms for DR staging from digital fundus images (DFIs) have been recently proposed. However, models often fail to generalize due to distribution shifts between the source domain on which the model was trained and the target domain where it is deployed. A common and particularly challenging shift is often encountered when the source- and target-domain supports do not fully overlap. In this research, we introduce DRStageNet, a deep learning model designed to mitigate this challenge. We used seven publicly available datasets, comprising a total of 93,534 DFIs that cover a variety of patient demographics, ethnicities, geographic origins and comorbidities. We fine-tune DINOv2, a pretrained model of self-supervised vision transformer, and implement a multi-source domain fine-tuning strategy to enhance generalization performance. We benchmark and demonstrate the superiority of our method to two state-of-the-art benchmarks, including a recently published foundation model. We adapted the grad-rollout method to our regression task in order to provide high-resolution explainability heatmaps. The error analysis showed that 59\% of the main errors had incorrect reference labels. DRStageNet is accessible at URL [upon acceptance of the manuscript].

\end{abstract}


%%%%%%%%%%%%%%%%%%%%
\section{Introduction}
\label{sec:introduction}

\IEEEPARstart{D}{iabetes} mellitus (DM) is one of the largest public health concerns globally \cite{Lin2020Global2025}. According to estimates by the International Diabetes Federation, 536.6 million people had DM in 2021, and prevalence is projected to increase to 783.2 million by 2045 \cite{idfAtlas}. Diabetic retinopathy (DR) is a direct microvascular end organ complication of DM. High glucose level caused by DM produces cytokines and growth factors that lead to capillary damage of eye blood vessels and causes increased vascular permeability and capillary occlusions. According to a 2012 study, approximately 34.6\% of DM patients suffer some degree of DR, 10. 2\% suffering from vision-threatening DR \cite{Yau2012GlobalRetinopathy}. Early detection of the disease is very important and any delay can result in rapid vision degradation and eventual irreversible blindness. Traditionally, DR is detected by a manual search for various lesions, including microaneurysms, hemorrhages, hard and soft exudates and vascular abnormalities \cite{Dubow2014ClassificationAngiography}. To avoid complications related to DR, patients with DM are recommended to undergo annual examinations \cite{Ting2017DevelopmentDiabetes}. The process requires highly skilled practitioners, with developing countries suffering from an acute shortage of such experts. Recently, deep learning (DL)-based algorithms for detection of DR from digital fundus images (DFI) have been suggested to tackle this challenge. Some of these algorithms focus on DR screening \cite{Gulshan2016DevelopmentPhotographs, Ting2017DevelopmentDiabetes, Voets2018ReproductionPhotographs, Abramoff2013AutomatedRetinopathy, Abramoff2016ImprovedLearning, Papadopoulos2021AnImages}, while others focus on DR staging \cite{Raiaan2023AImages, Vij2023AClassification, Martinez-Murcia2021DeepRetinopathy, Shaban2020ARetinopathy, Vijayan2023AEfficientnet}.

DR screening research and commercial devices consider the binary classification task of referable DR (rDR) \cite{Gulshan2016DevelopmentPhotographs, Ting2017DevelopmentDiabetes, Voets2018ReproductionPhotographs, Abramoff2013AutomatedRetinopathy, Abramoff2016ImprovedLearning, Papadopoulos2021AnImages}, which defines the positive class as a moderate or worse stage on the International Clinical Diabetic Retinopathy \cite{Wilkinson2003ProposedScales} (ICDR) scale or the presence of diabetic macular edema (DME). These models may be useful for nonophthalmologist professionals for the purpose of DR screening. DR staging research often models the task as a multiclass classification problem \cite{Raiaan2023AImages, Vij2023AClassification, Martinez-Murcia2021DeepRetinopathy, Shaban2020ARetinopathy}. Such models can support retina specialists in diagnosing DR and in monitoring disease progression and management. They can also be used by nonspecialists for screening. However, the main drawback of multi-class classification algorithms is the fact that the task is framed as a classification of multiple independent classes and does not take into account the ordinal relations between the classes.

The ability of DR staging models to generalize across diverse datasets remains a challenge. Models often fail to generalize \cite{Zhou2023AImages, Voets2018ReproductionPhotographs} due to distribution shifts between the source domain on which the model was trained and the target domain on which it is deployed. A common and particularly challenging shift often encountered in reality is where the source and target domain supports do not fully overlap. This commonly occurs with medical datasets that factor in differences in demographics, ethnicities, geographic origins, and/or comorbidities as well as technical specifications, in particular the type of camera and field of view (FOV). Finally, the explainability of DR staging algorithms is limited due to low causal relation between DR manifestations and the associated explainability heatmaps; i.e., there is a substantial amount of false positive and false negative regions that reduces their usability. This work makes the following contributions:
\begin{figure*}[ht]
    \centering
    \includegraphics[width=1\textwidth]{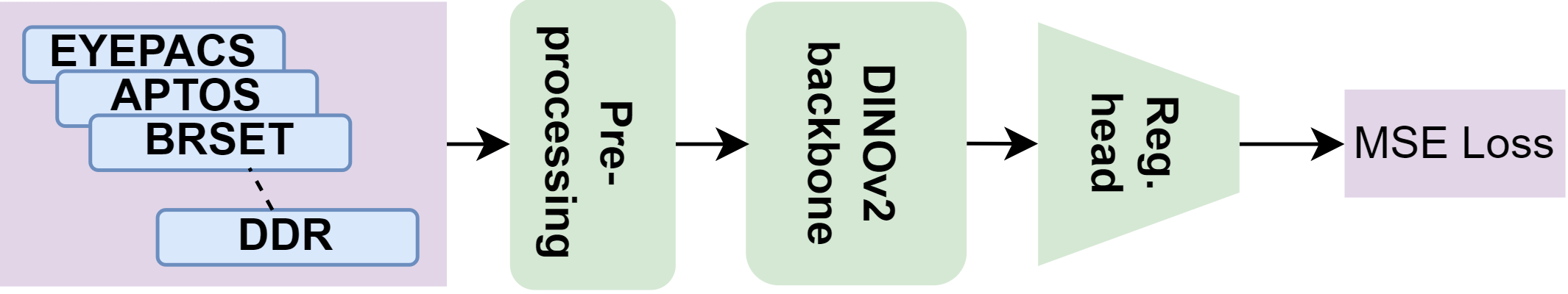}
    \caption{DRStageNet consists of a DINOv2 \cite{Oquab2023DINOv2:Supervision} pretrained backbone that is joined with a simple fully connected regression head. Additionally, we utilize a multi-source domain fine-tuning approach by combining seven open source DR datasets. DRStageNet is fine-tuned using the MSE loss.}
    \label{fig: architecture}
\end{figure*}
\begin{itemize}
    \item Introduction of DRStageNet (\cref{fig: architecture}), a robust, i.e., high-performing and generalizable, algorithm for DR screening, diagnosis, staging and progression monitoring.
    \item Adaptation of the grad-rollout \cite{jacobgildenblat} method to generate high-resolution explainability heatmaps for the regression task.
    \item Presentation of a detailed error analysis of DRStageNet in seven independent test sets.
\end{itemize}
%%%%%%%%%%%
\section{Datasets}
DR has multiple severity scales that are used in different countries and different clinics. The ICDR is the most common scale in open source DFI datasets \cite{Nakayama2021TheDataset}. For DR identification experiments, we selected seven independent open datasets (\cref{tab:datasets}) that had at least 500 DFIs available with ICDR grading. For each dataset, DFIs labeled nongradable, DFIs with missing labels, DFIs from children ($<18$ years) and DFIs from nondiabetes patients were excluded. rDR was defined using the DME labels when available and with the ICDR scale otherwise.

\subsubsection{Kaggle EYEPACS (Eye Picture Archive Communication System) \texorpdfstring{\cite{Dugas2015DiabeticKaggle.}}{Lg}}
The EYEPACS dataset was provided by the Eye Picture Archive Communication System \cite{Cuadros2009EyePACS:Screening} and was first introduced in the context of a Kaggle competition \cite{Dugas2015DiabeticKaggle.}. It contains 88,702 macula-centered DFIs of varying resolutions, which were captured by different cameras at different sites in the United States. The DFIs were classified for DR by a single clinician according to the ICDR scale using only the images as a reference. The dataset is divided into 35,126 DFIs used for training and 53,576 DFIs used for testing. Voets et al. \cite{Voets2018ReproductionPhotographs} found that approximately 20\% of the dataset are ungradable DFIs and redefined the dataset split to a training set of 28,132 DFIs from 14,404 patients (EYEPACS-train) and a test set of 42,922 DFIs from 24,524 patients (EYEPACS-test). We used this dataset and train-test split. For hyperparameters tuning, the training set was divided into a 90:10 split while stratifying at the patient level to avoid information leakage. The EYEPACS-test was used as the unseen source domain test set.

\subsubsection{DDR \texorpdfstring{\cite{Li2019DiagnosticScreening}}{Lg}}
The DDR dataset includes 13,673 macula-centered DFIs obtained from 9,598 patients from 147 hospitals in China. These images were captured using multiple cameras with a FOV of $45^{\circ}$. The dataset includes pixel-level and bounding box annotations of microaneurysms, hemorrhages and soft and hard exudates, as well as DR grades. Professional graders, who were trained by ophthalmologists, evaluated the DFIs using the ICDR scale on single-image level and also assessed their gradability. The final dataset consists of 12,519 DFIs.

\subsubsection{The Asia Pacific Tele-Ophthalmology Society (APTOS) \texorpdfstring{\cite{Karthik2019APTOSKaggle.}}{Lg}}
APTOS dataset contains a total of 5,590 macula-centered DFIs and it was made openly accessible by the Aravind Eye Hospital in India through a Kaggle competition. The images were captured by Aravind technicians in many rural regions of India, under varying conditions and environments and over a long period of time. The DFIs were later labeled by a group of doctors according to the ICDR scale without using additional information. The only accessible labels are of the APTOS-train split, which consists of 3,662 DFIs that we use in this research.

\subsubsection{Brazilian Multilabel Ophthalmological Dataset (BRSET) \texorpdfstring{\cite{NakayamaLuisFilipe2023APhysioNet}}{Lg}} 
This dataset comprises 16,266 DFIs centered on the macula, with a FOV of $45^{\circ}$, obtained from 8,524 patients examined at two ophthalmology clinics in Brazil (IRB 0698/2020). It includes demographic information, as well as anatomical parameters related to the macula, optic disc, and retinal blood vessels. Image quality parameters such as illumination, image field, and artifacts were also recorded to ensure quality control. The DFIs were graded on the single-image level using the ICDR scale. The data subset of diabetic patients consists of 1,301 individuals and 2,489 DFIs. This data subset was used for our experiments.

\subsubsection{Methods to Evaluate Segmentation and Indexing Techniques in the Field of Retinal Ophthalmology (MESSIDOR2) \texorpdfstring{\cite{Abramoff2013AutomatedRetinopathy}}{Lg}}
The Messidor-2 dataset is a collection of 1,748 macula-centered DFIs obtained using a Topcon TRC NW6 camera with $45^{\circ}$ FOV. These images are available in one of three resolutions: 1440 × 960, 2240 × 1488, or 2304 × 1536. While the dataset lacks official ICDR labels, alternative grading sets have been introduced by different research groups, all of which were based on a single DFI. Specifically, Google used a panel of 7 graders \cite{Gulshan2016DevelopmentPhotographs} who assessed the images using the ICDR scale. Labels provided by a different panel of 3 graders \cite{Krause2018GraderRetinopathy} were publicly released by Google. Additionally, the University of Iowa provided binary rDR grades\footnote{https://tinyurl.com/58sb5rm3}. For our research, the publicly accessible 3-grader annotations provided by Google \cite{Krause2018GraderRetinopathy} were used.

\subsubsection{The Indian Diabetic Retinopathy Dataset (IDRiD) \texorpdfstring{\cite{Porwal2018IndianIDRiD}}{Lg}}
IDRiD contains a total of 516 macula-centered DFIs that were acquired at an eye clinic located in Nanded, (M.S.), India. The DFIs were acquired with a Kowa VX-10 alpha with a $50^{\circ}$ FOV, and 4,288 × 2,848 pixels. The dataset contains pixel-level DR lesion and optical disc annotations, as well as image-based ICDR grade and binary classification of DME.

\subsubsection{Diabetic Retinopathy Two-field Image Dataset (DRTiD) \texorpdfstring{\cite{Hou2022Cross-FieldImages}}{Lg}} The DRTiD dataset contains a total of 3,100 two-field DFIs from 1,550 eyes, i.e., one macula-centered image and optical disc-centered imaged for each eye. Images were acquired between 2015 and 2017, using non-mydriatic retinal cameras with FOVs ranging between $45^{\circ}$ and $50^{\circ}$. The data acquisition was conducted as part of the Shanghai Diabetic Eye Study. A team of three experienced ophthalmologists graded the two-field DFIs using the ICDR scale. Intra-rater annotation discrepancies were reconciled by an expert ophthalmologist with clinical experience of more than 10 years. We used only the macula-centered DFIs for our experiments.

\begin{table*}[ht]
\centering
\caption{Description of the datasets used after removing DFIs that met the exclusion criteria. DR\% is the percentage of DR DFI in the dataset defined as an ICDR equal or superior to one. rDR\% is the percentage of rDR DFI in the dataset defined as an ICDR superior to one. NA denotes information unavailable and VAR denotes varying devices/fields of view.}
\label{tab:datasets}
\resizebox{2\columnwidth}{!}{%
\begin{tabular}{cccccccccc}
\textbf{Dataset} &
  \textbf{DFI} &
  \textbf{Patients} &
  \textbf{\begin{tabular}[c]{@{}c@{}}Device\end{tabular}} &
  \textbf{FOV} &
  \textbf{Resolution} &
  \textbf{Targets} &
  \textbf{DR\%} &
  \textbf{rDR\%} &
  \textbf{Origin} \\ \hline\hline
  
\begin{tabular}[c]{@{}c@{}}Kaggle\\ EYEPACS\end{tabular} &
  71,054 &
  40,529 &
  VAR &
  VAR &
  \begin{tabular}[c]{@{}c@{}}433 × 289 to\\  5184 × 3456\end{tabular} &
  ICDR &
  26 &
  19 &
  US \\ \hline
DDR &
  12,519 &
  9,598$<$ &
  VAR &
  45 &
  \begin{tabular}[c]{@{}c@{}}702 × 706 to\\ 5,184 3,456\end{tabular} &
  ICDR &
  50 &
  45 &
  China \\ \hline
APTOS &
  3,662 &
   NA &
  VAR &
  VAR &
  \begin{tabular}[c]{@{}c@{}}474 × 358 to\\ 4,288 2,848\end{tabular} &
  ICDR &
  51 &
  40 &
  India \\ \hline
BRSET &
  2,489 &
  1,301 &
  \begin{tabular}[c]{@{}c@{}}Nikon NF505\\ Canon CR-2\end{tabular} &
  45 &
  \begin{tabular}[c]{@{}c@{}}951 × 874 to\\ 2984 × 2304\end{tabular} &
  \begin{tabular}[c]{@{}c@{}}ICDR \\ DME \end{tabular} &
  26 &
  22 &
  Brazil \\ \hline
MESSIDOR2 &
  1,744 &
  874 &
  \begin{tabular}[c]{@{}c@{}}Topcon \\ TRC NW6\end{tabular} &
  45 &
  \begin{tabular}[c]{@{}c@{}}1440 × 960  to\\  2304 × 1536\end{tabular} &
  \begin{tabular}[c]{@{}c@{}}ICDR\\ DME \end{tabular} &
  42 &
  26 &
  France \\ \hline
DRTiD &
  1,550 &
   NA &
   NA &
  45-55 &
  \begin{tabular}[c]{@{}c@{}}1,444 × 1,444 to\\ 3,058 × 3,000\end{tabular} &
  \begin{tabular}[c]{@{}c@{}}ICDR \\ DME \end{tabular} &
  52 &
  42 &
  China \\ \hline
IDRiD &
  516 &
   NA &
  \begin{tabular}[c]{@{}c@{}}Kowa \\ VX-10a\end{tabular} &
  50 &
  4288 × 2848 &
  \begin{tabular}[c]{@{}c@{}}ICDR \\ DME \end{tabular} &
  67 &
  62 &
  India \\ %\hline
\end{tabular}%

}
\end{table*}

%%%%%%%%%%%%%%%%%%%%%%%%%%%%%%%%%%%%%%%%% 
\section{Methods}
\subsection{Deep learning for DR staging}

\subsubsection{Preprocessing}
First the horizontal black regions of the images are removed and then they are padded to a squared aspect ratio, based on the longest axis. The images were then resized to $518 \times 518$ pixels using bilinear interpolation. We found out that this preprocessing treatment obtained better performance than alternative techniques such as resizing without first cropping them to a square, or using Ben Graham's preprocessing technique \cite{Graham2015KaggleReport}, a preprocessing treatment which was used by other researchers developing DR algorithms \cite{Papadopoulos2021AnImages, Zhou2019CollaborativeImages, Zeng2019AutomatedNetwork}. 

\subsubsection{DRStageNet}
We approached the challenge of DR detection as a regression task against reference ICDR annotations \cite{Wilkinson2003ProposedScales}. The architecture (\cref{fig: architecture}) of DRStageNet consists of a pretrained DINOv2 \cite{Oquab2023DINOv2:Supervision} base vision transformer (ViT-base) \cite{DosovitskiyANSCALE}, having 86 million trainable parameters and a regression head that consists of two fully connected layers with a hidden dimension of 512 and a GeLU activation function \cite{Hendrycks2016GaussianGELUs}. Briefly, DINOv2 is a ViT that was trained on 142 million natural images using self-supervised learning (SSL). We chose to utilize transfer learning from SSL because it enables learning a representation based on a much larger dataset of natural images. This representation can then be fine-tuned for a specific task, i.e., subsequently training it using supervision. The output of the model is a single scalar and the total number of trainable parameters is 86.9 million. DRStageNet is fine-tuned using the mean squared error (MSE) loss function, where the target is the ICDR grade. The fine-tuning step is performed over the entire network, that is, without freezing any layer. We used a batch size of 16 DFIs with Adam \cite{Kingma2014Adam:Optimization} optimizer with a $0.04$ weight decay and a learning rate scheduler with an initial learning rate of $1e^{-6}$ which decreases 10-fold when the validation loss stopped decreasing over 4 epochs. Early stopping with respect to the validation loss was also used to reduce overfitting. Additionally, we used the data augmentations proposed by \cite{Voets2018ReproductionPhotographs}, consisting of random horizontal flips, jitter of contrast, saturation and hue. In our experiments, these augmentations proved to be the most stable and most effective compared to other ensembles of augmentations. We save the weights of the model with the lowest validation loss.

\subsubsection{Multi-source domain fine-tuning}
\begin{figure*}[htb]
    \centering
    \includegraphics[width=0.95\textwidth]{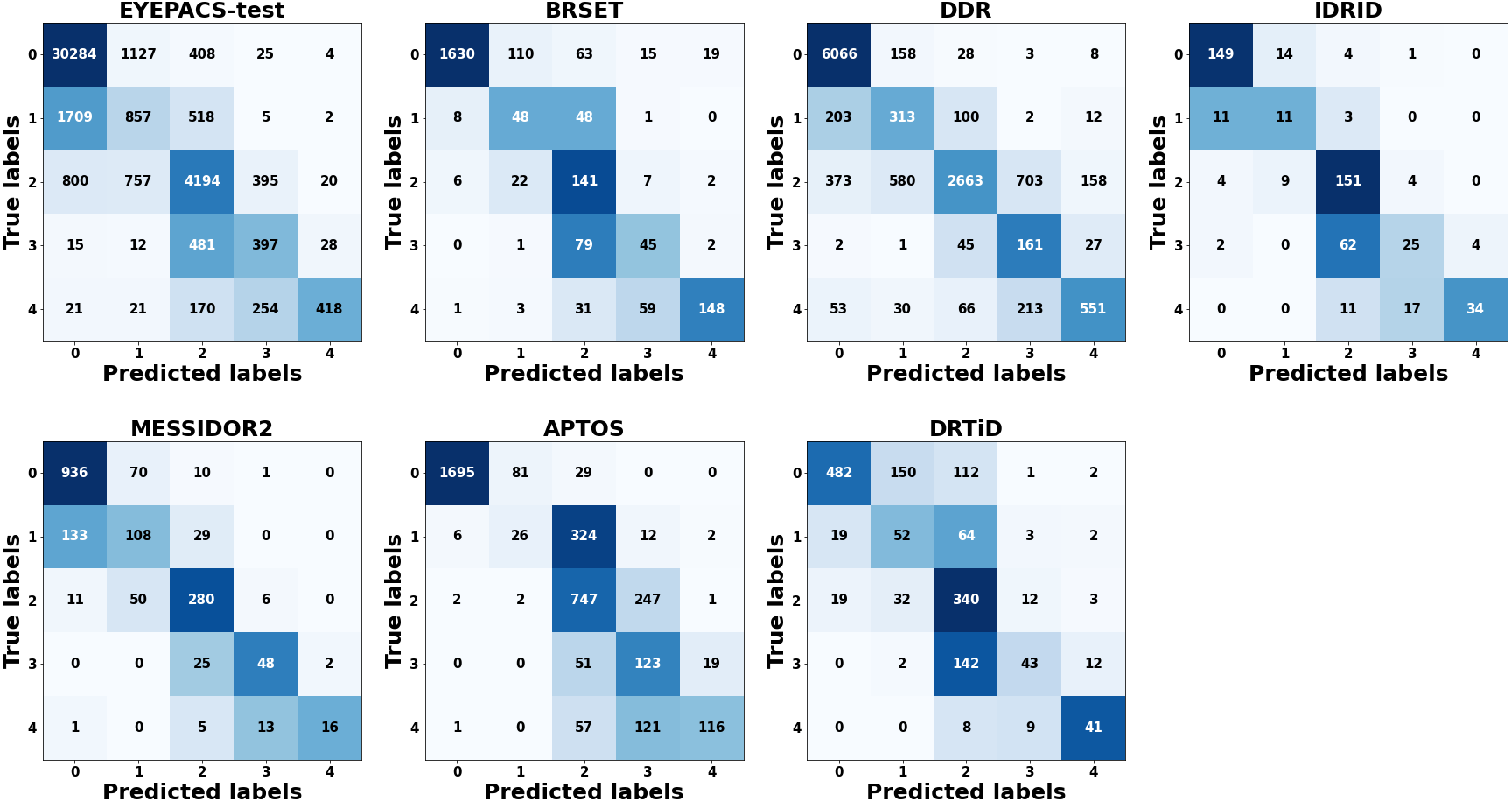}
    \caption{DRStageNet confusion matrices. For each confusion matrix but EYEPACS, the classification is reported for a model trained on all other datasets. For EYEPACS-test, the EYEPACS-train set and all other datasets are used to train the model.}
    \label{fig:DRStageNet confusion matrices}
\end{figure*}

We evaluated two methods for model training and evaluation. The first was a single-source domain (SS) fine-tuning on EYEPACS-train and evaluation on EYEPACS-test as well as on six external datasets (target domains). This method is called DRStageNet-SS. The second method uses multi-source domain fine-tuning (MST). It consists of training on a joint set of multiple source datasets while evaluating generalization performance on a single left-out target domain. The intuition behind this second approach is that when training a model on a single dataset, it may overfit this specific domain distribution. Variations in data collection equipment and inherent biases in the sample group, such as age, ethnicity, and health conditions, can cause such a model to fail when deployed. Furthermore, shortcut learning \cite{Geirhos2020ShortcutNetworks} can cause a model to recognize misleading patterns, leading to errors in real-world applications. An MST approach can moderate these effects by learning a wider support set as well as prevent the model from learning shortcut features. Training on multiple datasets should prevent shortcut features and overfitting of a specific population sample.
To implement MST, we split each of the seven datasets \cref{tab:datasets} for training and validation (90:10). The EYEPACS dataset was divided into EYEPACS-train which was included for all experiments and EYEPACS-test. At each fine-tuning stage, we used a joint validation dataset of all source domains aside from the left-out target domain.
To evaluate performance on a target domain, we used a leave-one-domain-out method, i.e., six out of seven datasets were used as source domains to train the model, while the left-out domain was used as the target domain. Therefore, in Figures \ref{fig: kappa_accuracy_mse_mae_subplots}, \ref{fig:DRStageNet confusion matrices}, \ref{fig: rdr auc and f1} and Tables \ref{tab:external-sota-comparison} and \ref{tab: DRStageNet results summary}, the performance measures reported are reported for EYEPACS-test, i.e., the test set of the source domain, while the other datasets were considered as unseen target domains. This MST approach is called DRStageNet.

\subsubsection{Benchmarks}
We benchmarked DRStageNet against two SOTA models \cite{Vijayan2023AEfficientnet,Zhou2023AImages}. Our first benchmark consisted of a large ImageNet \cite{Deng2010ImageNet:Database} pretrained EfficientNet2 \cite{Tan2021EfficientNetV2:Training} (117 million parameters). We added a regression head and fine-tuned it using the same protocol as DRStageNet-SS, using a larger initial learning rate of $1e^{-5}$. This model is similar to the work of Vijayan et al. \cite{Vijayan2023AEfficientnet}, with the one difference that we used EfficientNet2 \cite{Tan2021EfficientNetV2:Training}, which is the most contemporary version of EfficientNet. The second benchmark consisted of the fine-tuned version of the recently published pretrained  RETFound \cite{Zhou2023AImages} foundation model. RETFound is a ViT-large that consists of 303 million parameters, that was trained on a set of 1.6 million DFIs using the masked autoencoder \cite{He2021MaskedLearners} SSL method. Previous work showed that the use of transfer learning from a pretrained SSL model improved the performance of medical computer vision tasks \cite{Azizi2022RobustSelf-Supervision, Li2020Self-supervisedDiagnosis}, specifically DR diagnosis \cite{Truong2021HowTasks, Holmberg2020Self-supervisedRetinopathy, Burlina2022DetectingLearning}. We used the published source code\footnote{https://github.com/rmaphoh/RETFound\_MAE}. RETFound was fine-tuned using the same protocol as DRStageNet-SS while the initial learning rate was set to $1e^{-5}$ and the DFIs were resized to $224 \times 224$ pixels, which is the resolution input of RETFound.

%%%%%%%%
\subsection{Explainability}
A modified attention rollout \cite{Abnar2020QuantifyingTransformers} approach was used for model explainability. This method enables the use of the inherent high-resolution attention mechanism of our model, which is a transformer-encoder-based model. Briefly, the rollout \cite{Abnar2020QuantifyingTransformers} method describes how to compute the propagation of attention from the input to the last block of self-attention. Although this method has modest performance when used with images of a single large object, e.g., an image of a dog \cite{DosovitskiyANSCALE}, it tends to emphasize irrelevant tokens and is class-agnostic, i.e., the same heatmaps are attributed to different classes in the image. To address these issues, we followed the GradRollout \cite{jacobgildenblat} method and weighted the attention of each layer by its gradient with respect to the input. The GradRollout method was originally developed within the context of the classification framework in which gradient weights are taken with respect to the desired output class. In this research, we used a regression model such that the gradients were used with respect to the single-scalar output of the model. We hypothesized that, similarly to the classification setting, these gradients will emphasize the attribution of relevant information at each layer to a larger positive output. 

Let $x\in\mathbb{R}^{n\times n}$ be an input image and $f\left(x\right)\in\mathbb{R}$ be the output of the regression transformer model. We define $s-1$ to be the number of input tokens, which are the flattened patches across the image, $b$ to be the transformer self-attention block counter and $h$ be the number of heads in each transformer block. Then $A^{b}\in\mathbb{R}^{h\times \left(s\times s\right)}$ are the attention matrices at each block and $df/dA^{b} = W^{b}\in\mathbb{R}^{h\times \left(s\times s\right)}$ are the gradient weights that are multiplied element-wise with $A^{b}$. Our algorithm is described in equation \eqref{eq: grad-rollout}, where $A^{b+1}_{\text{gradrollout}}$ is the output heatmap at step $b+1$ and $g(\cdot)$ denotes first taking the maximum over the attention block heads, then zeroing out 10\% of the pixels of the lowest intensity. In addition,the attention matrix is normalized at the end of each step. After the algorithm reaches the last step of multiplication, $A^{b+1}_{\text{gradrollout}}$, it extracts the attention weights associated with the global classification token ([CLS] token) both horizontally and vertically and averages them, since the attention matrix is not necessarily symmetric (as a result of independent key and query matrices). Finally, we reshape the weights to a squared matrix and use a bilinear interpolation to resize them to the input dimension of $518\times518$ pixels.

\begin{align}
A_{\text{gradrollout}}^{b+1} &=\begin{cases}
A^{b} & ,b=0\\
\frac{1}{2}\left(g\left(A^{b}\odot W^{b}\right)+I\right)A_{\text{gradrollout}}^{b} & ,b>0
\end{cases} \label{eq: grad-rollout}
\end{align}

\subsection{Performance measures}
To assess the models' performance we used multiclass accuracy (MC-ACC), linearly weighted Cohen's kappa (LW-Kappa), MSE and mean absolute error (MAE) measures. We estimate the confidence interval by bootstrapping 1000 times 60\% of each test set, and the lower and upper bounds represent the quantiles of 0.25 and 0.75, respectively. Furthermore, we used the Mann-Whitney U statistical test on the performance of DRStageNet and a benchmark. We also report the area under the curve (AUC) metric for the binary rDR task. For that purpose, the DRStageNet output was transformed into a binary output, with the positive class defined as higher than stage one. The F1 score and the binary accuracy for the rDR task are also reported.

\subsection{Error analysis}
We examined instances where DRStageNet misclassified DFIs with a discrepancy of three or more units between the reference and predicted labels (\cref{fig:DRStageNet confusion matrices}). Each DFI underwent an independent and blinded review by two retinal specialists who were unaware of the dataset origin and the assigned reference label. The specialists assessed whether the DFIs were ungradable, identified the presence of one or more comorbidities, and assigned a DR grade according to the ICDR scale. Disagreements between the annotations of the two specialists were discussed, and a consensus was reached.

\begin{figure*}[htb]
    \centering
    \begin{minipage}[b]{0.47\linewidth}
        \centering
        \subfloat[]{\label{subfig: kappa plot}\includegraphics[scale=0.5]{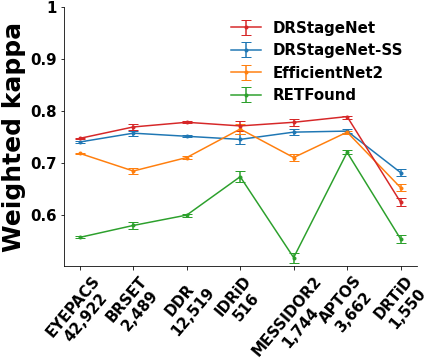}}
    \end{minipage} \hspace{0.005\linewidth}
    % \hfill
    \begin{minipage}[b]{0.47\linewidth}
        \centering
        \subfloat[]{\label{subfig: mc accuracy speed}\includegraphics[scale=0.5]{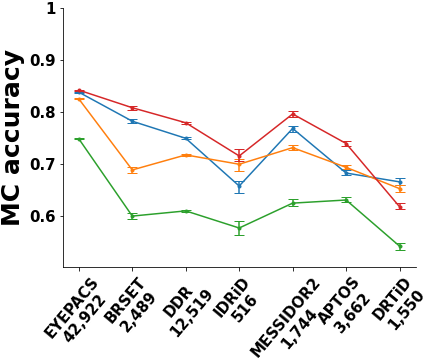}}
    \end{minipage}
    \begin{minipage}[]{0.47\linewidth}
        \centering
        \subfloat[]{\label{subfig: mse}\includegraphics[scale=0.5]{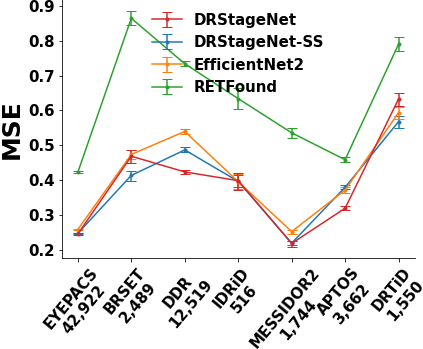}}
    \end{minipage}
    % \hfill
    \begin{minipage}[]{0.47\linewidth}
        \centering
        \subfloat[]{\label{subfig: mae}\includegraphics[scale=0.5]{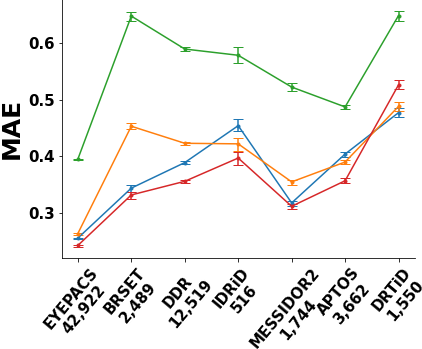}}
    \end{minipage}
    \caption{Models' DR staging performance across open-source datasets. (a) linearly weighted Cohen's kappa (LW-Kappa) performance. (b) Multiclass accuracy (MC-ACC) performance. (c) MSE performance. (d) MAE performance.}
    \label{fig: kappa_accuracy_mse_mae_subplots}
\end{figure*}

\begin{figure*}[htb]
    \centering
    \begin{minipage}[]{0.47\linewidth}
        \centering
        \subfloat[]{\label{subfig: rdr auc}\includegraphics[scale=0.5]{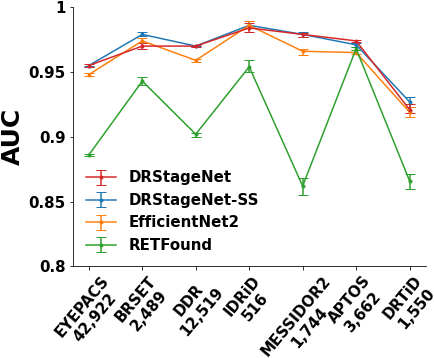}}
    \end{minipage}
    \begin{minipage}[]{0.47\linewidth}
        \centering
        \subfloat[]{\label{subfig: rdr f1}\includegraphics[scale=0.5]{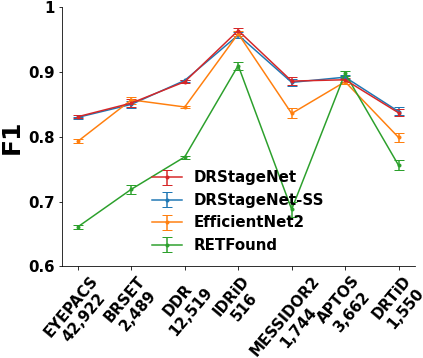}}
    \end{minipage}
    \caption{Models' rDR performance across open-source datasets. (a) rDR AUC performance. (b) rDR F1 performance.}
    \label{fig: rdr auc and f1}
\end{figure*}

\begin{figure*}[!t]
    \centering
    \includegraphics[width=0.9\linewidth]{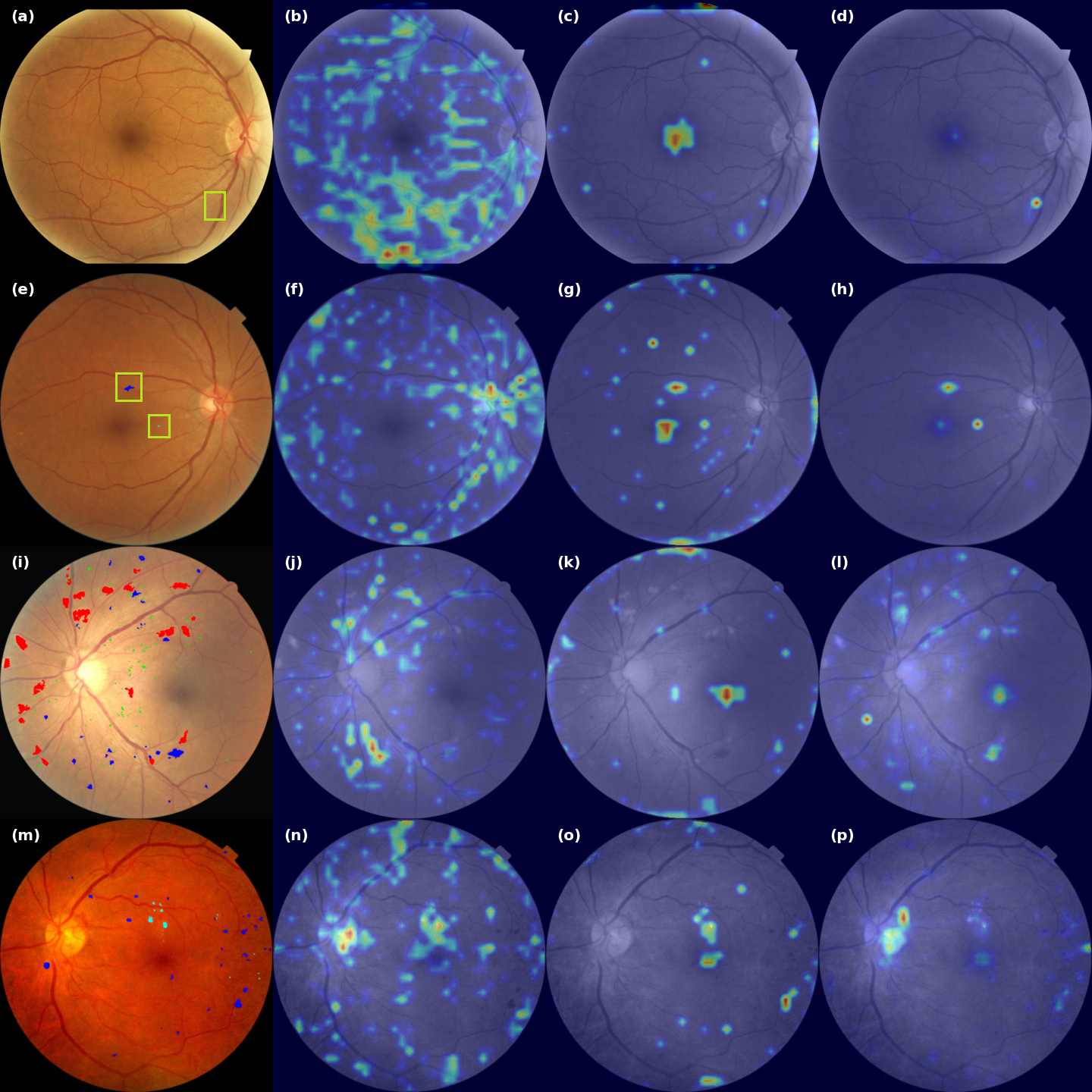}
    \caption{A comparison between three explainability methods. The first column represents a set of four DFIs from the DDR dataset. They were ordered by their respective ICDR reference labels. Image (a) is labeled mild diabetic retinopathy (stage 1) while image (m) is labeled proliferative diabetic retinopathy (stage 4). The ground-truth microaneurysms, hemorrhages, soft exudates and hard exudates were colored in green, blue, red, and cyan, respectively. The ground-truths in images (a) and (e) have bounding boxes for visual support. The second column shows the heatmap output of Grad-CAM \cite{Selvaraju2016Grad-CAM:Localization} method for the last transformer block of DRStageNet, the third column shows the heatmaps from the rollout \cite{Abnar2020QuantifyingTransformers} method and the fourth column is our explainability heatmaps. In image (p), DRStageNet attends to some neovascularization near the optic disc that is not present in the ground-truth segmentation.}
    \label{fig: explainability}
\end{figure*}

%%%%%%%%%%%%%%%%%%%%%%%%%%%%%%%%%%%

\begin{table}[htb]
\caption{External performance comparison of rDR AUC and MC-ACC between DRStageNet and other methods. (*) The reported results are estimated from Fig. 2b in the paper \cite{Zhou2023AImages}.}
\label{tab:external-sota-comparison}
\resizebox{\columnwidth}{!}{%
\begin{tabular}{c|cccccc}

\textbf{\begin{tabular}[c]{@{}c@{}}Eval. dataset\end{tabular}} &
  \textbf{Method} &
  \textbf{rDR AUC} &
  \textbf{MC-ACC} \\ \hline\hline

\multicolumn{1}{c|}{APTOS}     & \textbf{DRStageNet}                                     & \textbf{0.974} & \textbf{0.739} \\  
\multicolumn{1}{c|}{}          & DRGen  \cite{Atwany2022DRGen:Classification}        &                & 0.703        \\
\multicolumn{1}{c|}{}          & RETFound \cite{Zhou2023AImages}                   & 0.800*          &       \\ \hline
\multicolumn{1}{c|}{MESSIDOR2} & \textbf{DRStageNet}                                     & \textbf{0.979} & \textbf{0.796} \\  
\multicolumn{1}{c|}{}          & Papadopoulos et al. \cite{Papadopoulos2021AnImages}       & 0.976          &              \\  
\multicolumn{1}{c|}{}          & DRGen \cite{Atwany2022DRGen:Classification}         &                & 0.705        \\ 
\multicolumn{1}{c|}{}          & RETFound \cite{Zhou2023AImages}                   & 0.820*          &       \\ \hline
\multicolumn{1}{c|}{IDRiD}     & \textbf{DRStageNet}                                     & \textbf{0.984} & \textbf{0.717} \\
\multicolumn{1}{c|}{}          & RETFound \cite{Zhou2023AImages}                   & 0.820*          &       \\ 

\end{tabular}%
}
\end{table}

\section{Results}
\subsection{Generalization performance}
Figure \ref{fig: kappa_accuracy_mse_mae_subplots} presents the measured LW-Kappa, MC-ACC, MSE and MAE for DRStageNet, DRStageNet-SS and two contemporary benchmarks across the seven datasets used in this research (\cref{tab:datasets}). DRStageNet LW-Kappa on the local EYEPACS test set was 0.747 vs. 0.718 for the EfficientNet2 SOTA benchmark. The generalization performance of DRStageNet was significant (p<0.001) and non-incremental over the EfficientNet2 SOTA benchmark for five out of six of the external datasets. The only exception was observed in the MSE metric on the IDRID dataset, where DRStageNet and EfficientNet2 exhibited similar performance. The confusion matrices for DRStageNet are displayed in \cref{fig:DRStageNet confusion matrices}. It can be observed that most errors were one step off the diagonal, which is similar to human inter- and intra-rater inconsistency \cite{Atwany2022DeepSurvey}.

In addition, \cref{tab:external-sota-comparison} compares DRStageNet to other works reporting on the performance on one or more of the external datasets included in our study. DRStageNet demonstrated superior performance in terms of MC-ACC compared to DRGen \cite{Atwany2022DRGen:Classification}, which used a domain generalization approach to train their model. Furthermore, on the MESSIDOR2 dataset, DRStageNet achieved an rDR AUC score of 0.979, similar to the performance of Papadopoulos et al. \cite{Papadopoulos2021AnImages}. Additionally, the reported rDR AUC performance of RETFound \cite{Zhou2023AImages} was lower than DRStageNet's rDR AUC scores, e.g., 0.82 vs. 0.979 AUC on MESSIDOR2. 

We report the rDR AUC performance of DRStageNet versus the other two benchmarks (\cref{fig: rdr auc and f1}). DRStageNet, DRStageNet-SS and EfficientNet2 showed comparable performance on all the datasets. RETFound exhibited a lower performance than the other three methods, but it had a higher performance than previously reported in the author's original article \cite{Zhou2023AImages}, e.g., 0.968 versus approximately 0.8 AUC on APTOS. A similar performance comparison of rDR F1, MSE and MAE and detailed numerical results for DRStageNet is presented in \cref{fig: rdr auc and f1}, \cref{fig: kappa_accuracy_mse_mae_subplots} and \cref{tab: DRStageNet results summary} respectively.

\begin{table}[htb]
\caption{DRStageNet performance summary on the datasets used in this research. The reported kappa is linearly weighted and ACC is the binary rDR accuracy.}
\label{tab: DRStageNet results summary}
\resizebox{\columnwidth}{!}{%
\begin{tabular}{c|cccccc}

\textbf{\begin{tabular}[c]{@{}c@{}}Eval. dataset\end{tabular}} &
  \textbf{AUC} &
  \textbf{ACC} &
  \textbf{F1} &
  \textbf{MC-ACC} &
  \textbf{\begin{tabular}[c]{@{}c@{}}Kappa\end{tabular}} &
  \textbf{MSE} \\ \hline\hline
\multicolumn{1}{c|}{EYEPACS}   & 0.955 & 0.940 & 0.831 & 0.842 & 0.747 & 0.246\\
\multicolumn{1}{c|}{DDR}       & 0.970 & 0.905 & 0.885 & 0.779 & 0.779 & 0.424\\ 
\multicolumn{1}{c|}{APTOS}     & 0.974 & 0.898 & 0.888 & 0.739 & 0.789 & 0.320\\ 
\multicolumn{1}{c|}{BRSET}     & 0.970 & 0.928 & 0.852 & 0.808 & 0.769 & 0.468\\
\multicolumn{1}{c|}{MESSIDOR2} & 0.979 & 0.942 & 0.886 & 0.796 & 0.777 & 0.218\\
\multicolumn{1}{c|}{DRTiD}     & 0.922 & 0.847 & 0.837 & 0.618 & 0.624 & 0.633\\
\multicolumn{1}{c|}{IDRiD}     & 0.984 & 0.955 & 0.964 & 0.717 & 0.772 & 0.397

\end{tabular}%
}
\end{table}

\subsection{Explainability}

We validated the DRStageNet heatmaps by comparing them with the ground truth DR lesion annotations of the DDR dataset (\cref{fig: explainability}) of four DFIs that correspond to the four stages of DR. The heatmaps correctly highlight the annotated lesions in the four DFIs. In the DFI heatmap that corresponded to a proliferative case (p), DRStageNet highlighted some neovascularization, i.e., formation of new blood vessels, near the optic disc, which were not annotated in the ground truth. A DFI is classified as proliferative DR when there exists neovascularization or vitreous/preretinal hemorrhage. In this case, DRStageNet attended the neovascularization near the optic disc, which may explain its correct classification. Furthermore, \cref{fig: explainability} presents a comparative analysis of our explainability approach with Grad-CAM and Rollout methods. Qualitatively, we appreciate on this set of examples that these techniques generate an important number of false positives. This makes them less valuable in providing an explainability support tool to a prospective clinical user.

\subsection{Error analysis}
A total of 247 DFIs were reviewed. Among these, a total of 106 (43\%) DFIs had at least one comorbidity, 24 (10\%) were considered ungradable, and 147 (59\%) were mislabelled. Overall, a total of 210 (85\%) DFIs were either ungradable, had a comorbidity or were mislabelled.

\section{Discussion}
\begin{figure*}[htb]
    \centering
    \begin{minipage}[]{0.3\linewidth}
        \centering
        \subfloat[]{\label{subfig: mislabelling}\includegraphics[width=\linewidth]{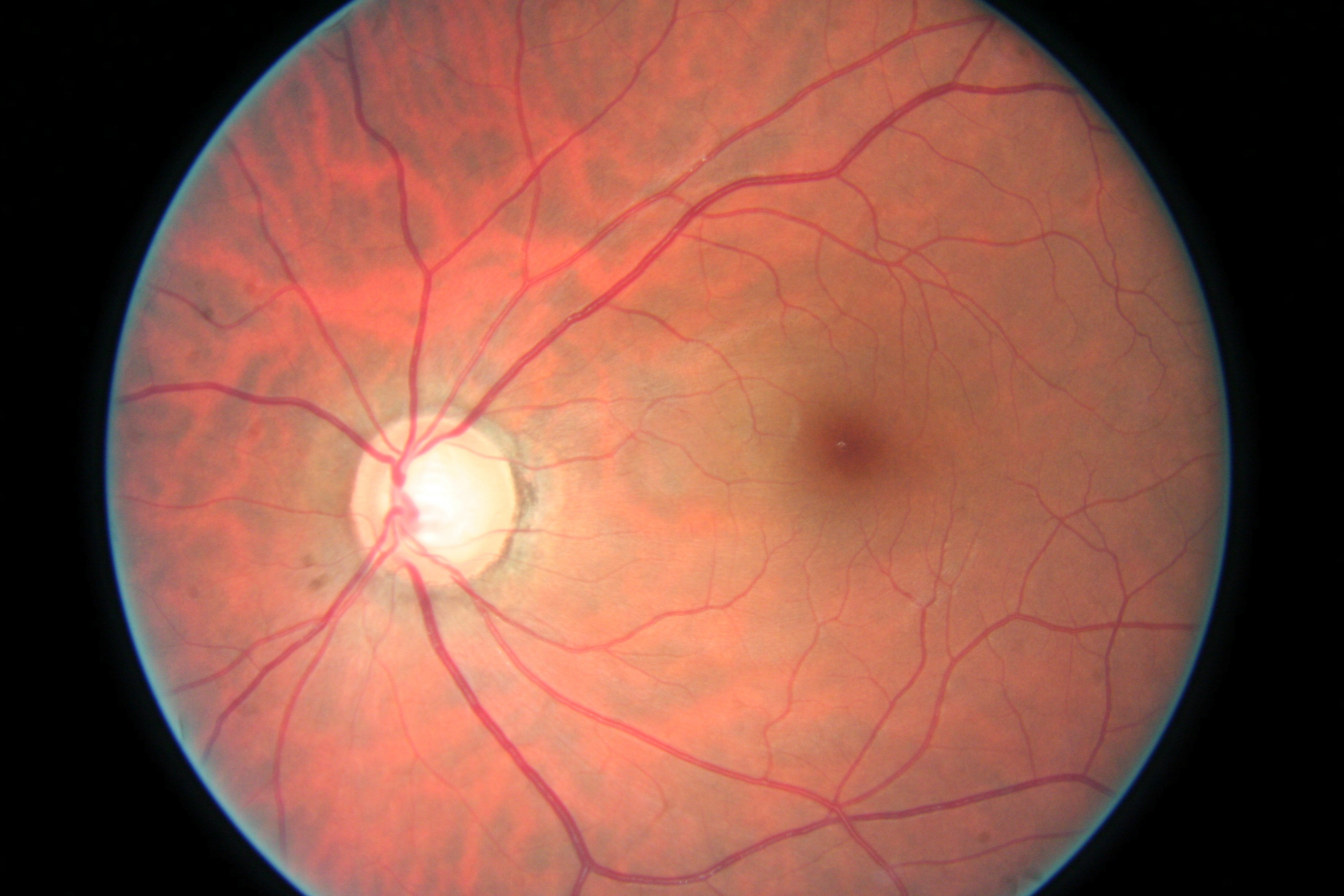}}
    \end{minipage}
    % \hfill
    \begin{minipage}[]{0.3\linewidth}
        \centering
        \subfloat[]{\label{subfig: comorbidity}\includegraphics[width=\linewidth]{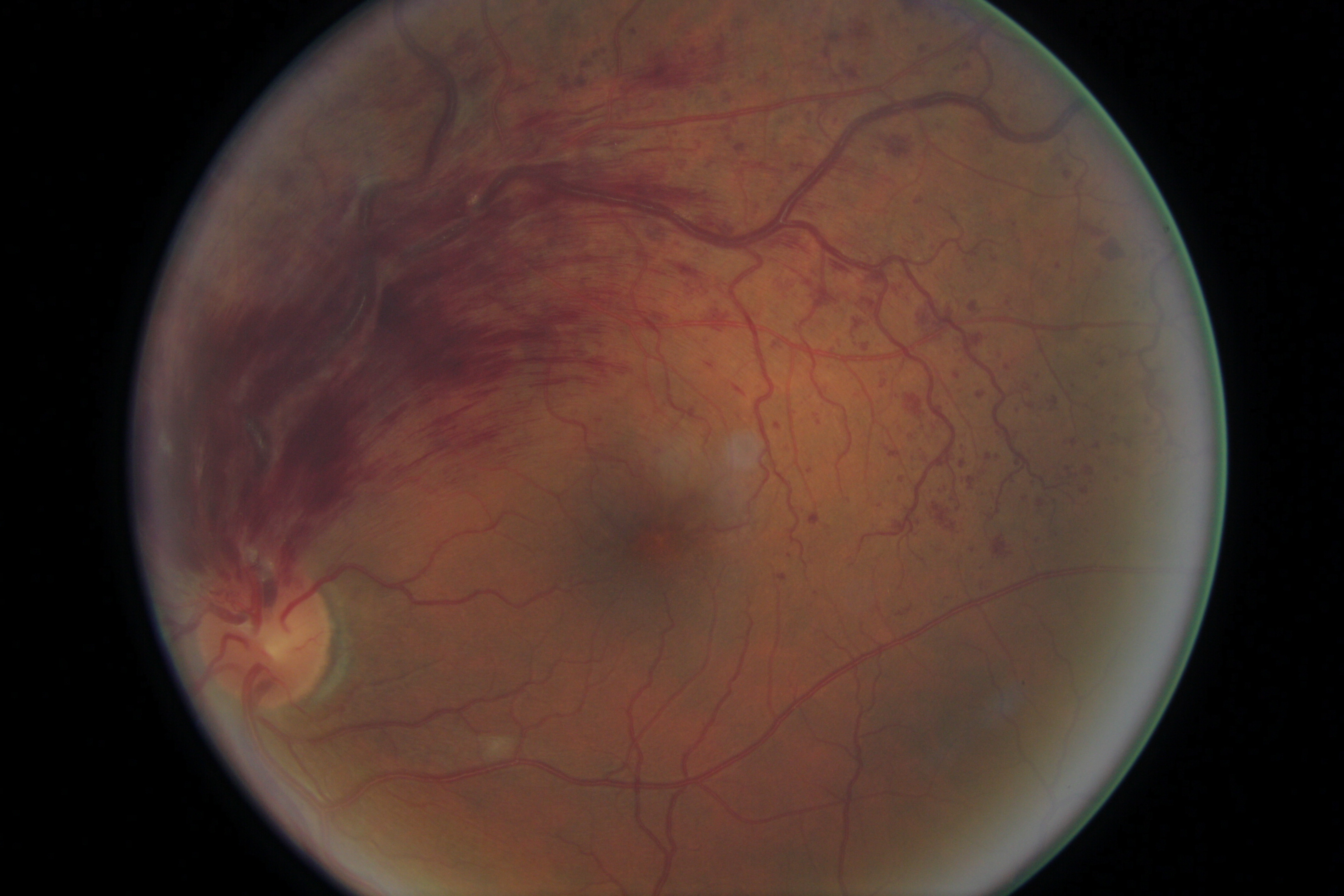}}
    \end{minipage}
    % \hfill
    \begin{minipage}[]{0.268\linewidth}
        \centering
        \subfloat[]{\label{subfig: ungradable}\includegraphics[width=\linewidth]{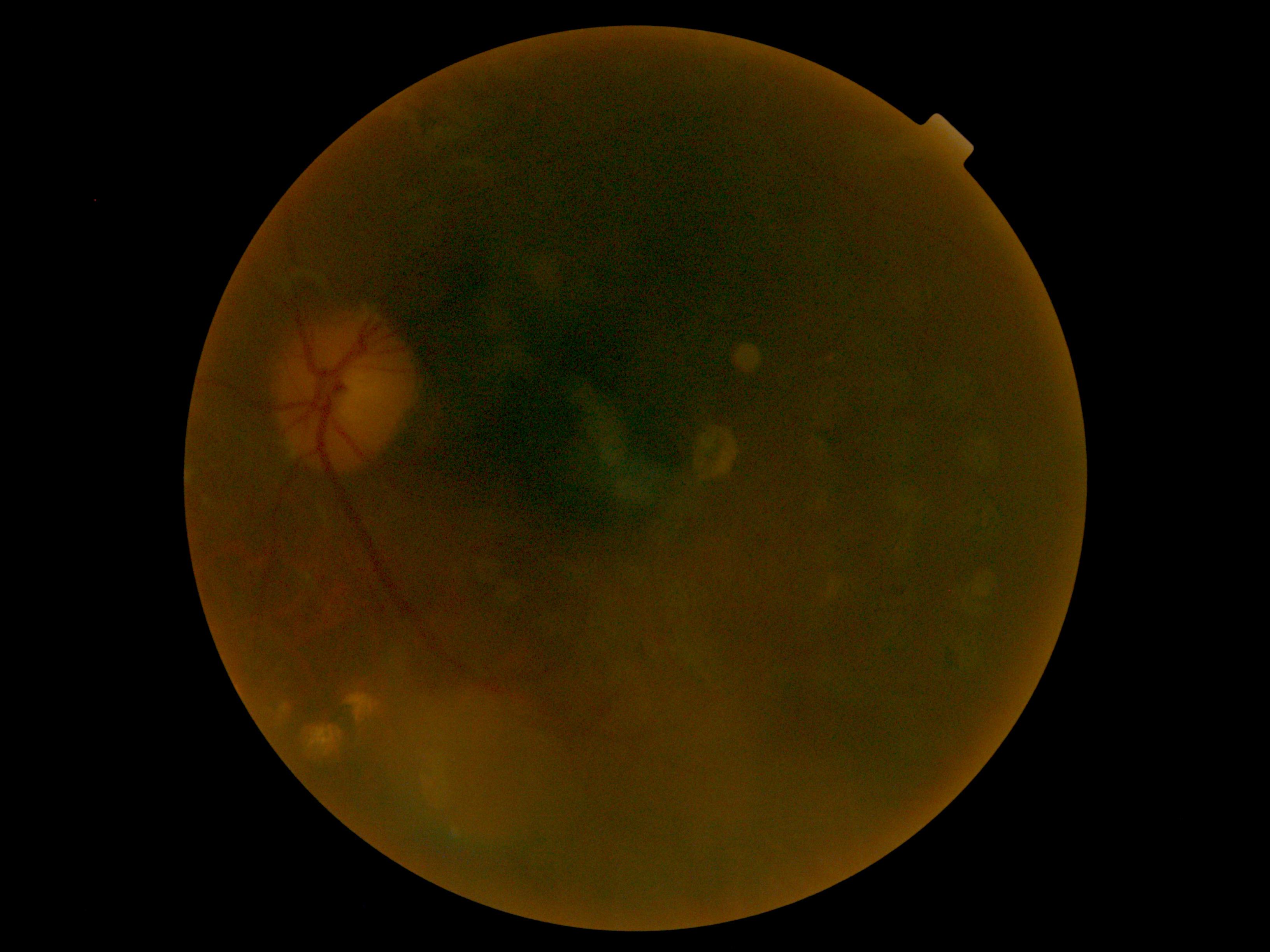}}
    \end{minipage}
    \caption{Some examples of the extreme errors of DRStageNet. (a) Mislabelling example from EYEPACS dataset. Original ICDR target is 4 and the reviewed target 0. DRStageNet predicted stage 0 correctly. (b) A DFI with a comorbidity example from EYEPACS dataset, specifically of vascular occlusion, which might resemble multiple hemorrhages and microaneurysms similar to DR stage 3 (severe non-proliferative DR), as predicted by DRStageNet. (c) Ungradable DFI example from DDR dataset. This DFI has lighting and focus problems.}
    \label{fig: mislabelling examples}
\end{figure*}

The single-source DRStageNet-SS method outperformed the other two single-source benchmarks on the source domain EYEPACS-test set, and exhibited equal or superior generalization performance on all target domains except for IDRiD. These results demonstrate the value of using a transformer architecture that was pretrained using SSL on a large number of natural images to create a representation that can be fine-tuned to a specific downstream classification task. Our MST approach, DRStageNet, exhibited even better generalization performance except for on the DRTiD dataset, on which all the methods showed reduced performance. The results obtained for the MST DRStageNet algorithm highlight the value of using MST in learning a more generalizable representation for a given task, as it avoids overfitting a specific domain or learning shortcut features. The confusion matrices for DRStageNet (\cref{fig:DRStageNet confusion matrices}) show that the majority of errors are one step off the diagonal. This is similar to human inter- and intra-rater inconsistencies \cite{Atwany2022DeepSurvey}. For the secondary task rDR, performance was superior but incremental compared to the EfficientNet2 benchmark. Indeed, the binary task is simpler than ICDR staging and does not require the global attention mechanism that distinguishes transformer models such as the one used in DRStageNet.

The generalization performance of all algorithms was low for the DRTiD dataset. In DRTiD, the ground truths are based on two-field DFIs, one being macula-centered while the other is optic disc-centered. Therefore, unlike the other datasets used, the ICDR label is provided at the patient level as opposed to the image level. These differences in the results obtained with a dataset annotated at the patient level versus the single DFI level suggest the need for integration of multiple DFIs of the retina, at least one macula-centered and one disc-centered.

Our error analysis revealed that the majority of misclassified DFIs with a gap of three or more were incorrectly labeled (63\%). This is a recognized issue in open-source DFI datasets, which has prompted some researchers \cite{Gulshan2016DevelopmentPhotographs, Krause2018GraderRetinopathy} to re-annotate datasets by engaging multiple DR experts to refine the reference labels. Additionally, a considerable number of misclassified DFIs were associated with at least one comorbidity (40\%). In fact, certain comorbidities can be confused with DR due similarities in their patterns, e.g., vascular occlusion (see \cref{fig: mislabelling examples}), or because they are relatively rare pathologies that are not adequately represented in the training set, preventing the network from learning a meaningful representation.

Our explainability heatmaps exhibited high association with ground-truth lesion segmentation of the DDR dataset. However, there were still some discrepancies. The heatmaps sometime highlighted regions such as the macula or the optic disc that suggest that DRStageNet identifies some important features in these regions. Some DR manifestations were, however, not highlighted, e.g., in \cref{fig: explainability} (i). It is possible that DRStageNet's decision may be driven by a subset of DR objects while not attending the others. The main goal of the heatmaps is to help specialists identify DR patients, which DRStageNet helps achieve. We believe that the combination of the ICDR regression outcome and the attention heatmaps will help reduce clinician workload and the number of false negatives cases of DR patients.

In this research, we introduced DRStageNet, a DL model for DR, which aims to accurately stage DR and mitigate the challenge of generalizing performance to target domains. For this purpose, DRStageNet uses a SSL-based pretrained ViT model and implements a multi-source domain fine-tuning strategy. We demonstrated the superiority of this method over two SOTA models.

\printbibliography

\end{document}